\documentclass[aps,pra,preprint,groupedaddress]{revtex4-2}
\usepackage{graphicx,caption,subcaption}
\newcommand{\eqref}[1]{(\ref{#1})}%refer to equations
\renewcommand{\varphi}{\wp} % define \varphi to be \wp (Weierstrass rho)
\begin{document}

%\preprint{}

\title{Comment on "Rogue waves on the double-periodic background in the focusing nonlinear Schrödinger equation"}


\author{Hans Werner Sch\"urmann}
\email[]{hwschuer@uos.de}
\affiliation{Department of Mathematics, Computer Science, and Physics\\ University of Osnabr\"uck, Germany}

\author{Valery Serov}
\email[]{vserov@cc.oulu.fi,valserov@gmail.com}
\affiliation{Research Unit  of Mathematical Sciences\\ University of Oulu, Finland}
%Moscow Centre of Fundamental and Applied Mathematics-\\ - Lomonosov Moscow State University, Russia}


%\date{}

\begin{abstract}
Some years ago, Chen, Pelinovsky, and White claimed existence of certain solutions of the nonlinear Schr\"odinger equation for modelling rogue waves [arXiv: 1909.08165v1 (2019)]. It is the aim of this Comment to outline that this claim is faulty. -- The solutions presented do not satisfy the nonlinear Schr\"odinger equation.
\end{abstract}

%\keywords{}

\maketitle

%\section{}
%\subsection{}
%\subsubsection{}

%\section{Introduction}

In an article \cite{ChPW}, Chen, Pelinovsky, and White presented solutions (Eqs.(1.2), (1.3) in \cite{ChPW}) of the nonlinear Schr\"odinger equation (NLSE)
\begin{equation}
i\Psi_t(x,t)+\frac{1}{2}\Psi_{xx}(x,t)+\Psi(x,t)|\Psi(x,t)|^2=0
\end{equation}
by using the ansatz ((3.8) in \cite{ChPW})
\begin{equation}
\Psi(x,t)=(Q(x,t)+i\delta(t))e^{i\theta(t)}.
\end{equation} 
Substituting (2) into (1) and separating real and imaginary parts they derived the system of partial differential equations -- that must be valid necessarily if ansatz (2) is compatible with the NLSE (1) --
\begin{equation}
Q_{xx}(x,t)+2(Q^2(x,t)+\delta^2(t)-\theta_t(t))Q(x,t)-2\delta_t(t)=0
\end{equation}
\begin{equation}
Q_t(x,t)+(Q^2(x,t)+\delta^2(t)-\theta_t(t))\delta(t)=0,
\end{equation}
and evaluated (3), 
%using (partly) results of \cite{AEK}, 
to obtain solutions for $\delta(t)$ and $Q(x,t)$ (Eqs.(3.24),(3.28),(3.29),(3.32),(3.38) in \cite{ChPW}). Finally they derived solutions (1.2) (from (3.23), (3.28)) and (1.3) (from (3.32), (3.38)).

In a recent article \cite{SchSe} we considered the nonlinear Schr\"odinger equation and solution ansatz in the form
\begin{equation}
i\Psi_z(t, z)+\Psi_{tt}(t, z)+a\Psi(t, z)|\Psi(t, z)|^2=0
\end{equation}
\begin{equation}
\Psi_z(t, z)=(f(t,z)+id(z))e^{i\Phi(z)}
\end{equation}
with the assumptions $d_z(z)\ne 0, f_t(t,z)\ne 0$.

With $a=2, \frac{z}{2}\to z,$ and apart from different notations of the variables and of the ansatz function, system $\{(5a), (5b)\}$ in \cite{SchSe} is identical with system $\{(3), (4)\}$. -- Hence, results of \cite{SchSe} can be applied to $\{(3), (4)\}$. 

In \cite{SchSe}, we transformed system $\{(5a), (5b)\}$ to the dynamical system $\{(14a), (14b)\}$ that, in the notations of \cite{ChPW} reads
\begin{equation}
Q_t(x,t)=\sqrt{z(t)}(c_1-2(3z(t)+Q^2(x,t)))
\end{equation}
\begin{equation}
(Q_x(x,t))^2=-Q^4(x,t)+(c_1-6z(t))Q^2(x,t)+\frac{z_t(t)}{\sqrt{z(t)}}Q(x,t)+2c_2+3z^2(t)-c_1z(t)
:= R_2(Q(x,t), t)
\end{equation}
where $z(t)=\delta^2(t)$, and $c_1,c_2,c_3$ are free integration constants (see Eqs.(8a)-(8c) in \cite{SchSe}). Function $z(t)=\delta^2(t)$ is uniquely determined by (see (5) in \cite{SchSe} with $q=2$)
\begin{equation}
(z_t(t))^2=-64z^4(t)+32c_1z^3(t)-4(c_1^2+8c_2)z^2(t)+4c_3z(t)=:R_1(z).
\end{equation}
Function $Q(x,t)$ is uniquely determined by Eq.(8). Explicit expressions for $z(t)$ and $Q(x,t)$ are given by Eqs.(7) and (8) (with $q=2$) in \cite{
SchSe-II}, respectively. Functions $z(t)$ and $Q(x,t)$, as the solutions of Eq.(9) and Eq.(8), respectively, are valid for any parameters $c_1, c_2, c_3, z_0, Q_0$ and variables $x,t$. We consider real parameters and appropriate $z_0$ and $Q_0$ and real $x$ and $t$, so that the invariants of Weierstrass' functions $\wp$ are real and thus $z(t)$ and $Q(x,t)$, since $\wp$ is real for real arguments and real invariants. Evaluation of $z(t)$ according to (7) in \cite{SchSe-II} yields (with $z_0=0$)
\begin{equation}
z(t)=\frac{3c_3}{c_1^2+8c_2+\wp(t; g_{2z},g_{3z})},
\end{equation}
where
$$
g_{2z}=\frac{4(c_1^2+8c_2)^2}{3}-32c_1c_3,\quad g_{3z}\frac{8}{27}\left((c_1^2+8c_2)^3-36c_1c_3(c_1^2+8c_2)+216c_3^2\right).
$$
Equation (8) in \cite{SchSe-II} can be evaluated to (with $q=2$):
$$
Q(x,t; Q_0)=
$$
\begin{equation}
\frac{-2\gamma_2\delta_2 - (5\gamma_2^2-\alpha_2\epsilon_2)Q_0 + 2\alpha_2\delta_2 Q_0) + 4\wp(x)(\delta_2+2\gamma_2Q_0+\wp(x)Q_0) + 
2\wp_x(x)\sqrt{R_2(Q_0, t)}}{(2\wp(x)-\gamma_2-\alpha_2Q_0^2)^2-\alpha_2R_2(Q_0,t)}
\end{equation}
with $\wp(x)=\wp(x; g_{2Q}, g_{3Q}), \alpha_2=-1,\gamma_2=\frac{1}{6}(c_1-6z(t)), \delta_2=\frac{z_t(t)}{4\sqrt{z(t)}},\epsilon_2=\frac{z(t)}{2}(6z(t)-2c_1)+2c_2$, and with
$$
g_{2Q}=\frac{c_1^2}{12}-2c_2,\quad g_{3Q}=-\frac{c_1^3}{216}-\frac{c_1c_2}{3}+\frac{c_3}{4}.
$$
System $\{(7), (8)\}$ is necessarily valid, if system $\{(3), (4)\}$ ((3.14) in \cite{ChPW}) is assumed to be valid. Function $Q(x,t)$, according to (11), is the only possible solution of system $\{(7), (8)\}$, and hence of $\{(3), (4)\}$. Thus the validity of (3.14) in \cite{ChPW} must be checked only with (11) substituted into Eqs.(8) and (7). Straightforward analytical evaluation shows that Eq.(8) is satisfied by (11). This does not hold for Eq.(7) with Eq.(11) substituted due to numerical evaluation (see Fig.1). Thus system (3.14) in \cite{ChPW} is inconsistent.

We note, this result does not imply that solutions $z(t)$ and $Q(x,t)$ presented in \cite{ChPW} are "wrong" in a certain sense. As far as we can see, they are -- consistent with (10) and (11) -- correct solutions of the "first-order quadratures" Eqs.(3.12) and (3.15), and hence are correct solutions only of the first equation, not of the second equation of (3.14). It would have been appropriate if the authors would have checked both equations with $z(t)$ and $Q(x,t)$ according to (3.23), (3.28) and (3.32), (3.38) in \cite{ChPW}.

To conclude, we have presented arguments for the inconsistency (if $\delta_t(t)\ne 0, Q_x(x,t)\ne 0$) of system $\{(1), (2)\}$ in \cite{ChPW}. The claim in Section V "This work opens up a number of new directions in the study of rogue waves modelled by the focusing NLS equation" is (at least) doubtful, though it has found more than 80 citations in the Web of Science.

%FIG1
\begin{figure}[h!]
\includegraphics[width=14cm]{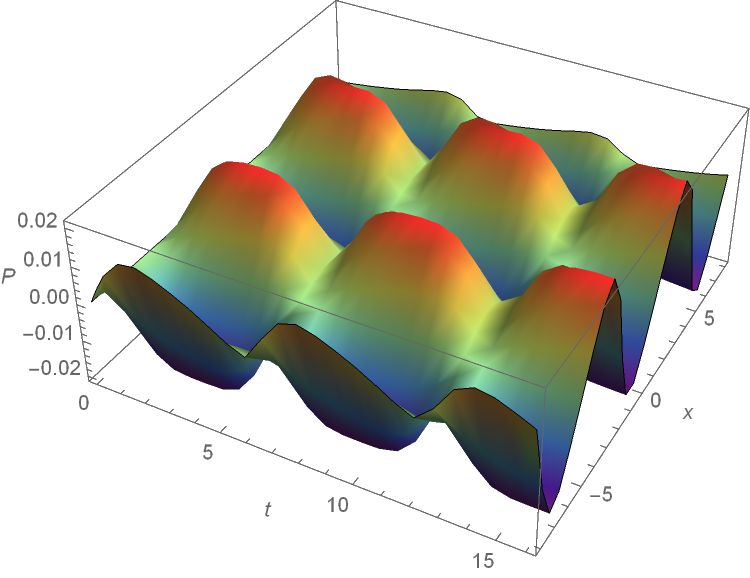}
\caption{$P=Q_t(x,t)-\sqrt{z(t)}(c_1-6z(t))-2Q^2(x,t)$ for $z_0=0$ $Q_0=\frac{1}{2}, c_1=\frac{3}{4}, c_2=-\frac{3}{128}, c_3=\frac{1}{256}, q=2$.}
\end{figure}

\section*{note:}

The foregoing Comment has been submitted as a Comment on the PRE-version of \cite{ChPW} (Phys.Rev.E 100, 052219 (2019)) to PRE and been rejected.

From our point of view, we have disproved the objections of the Referees in the review process. Unfortunately, it is not clear whether we have permissions to reproduce the details of the peer review process in the present Comment. What we can state is: None of the reviewers has refereed to our main claim that Eq.(7) is not satisfied by $z(t), Q(x,t,Q_0)$ according to (10), (11), respectively (connected to the doubts on the validity of (3.14) in \cite{ChPW}).

To conclude, we do not doubt the intellectual honesty of the reviewers, but it belongs to the virtues and duties of Science to indicate errors if they are noticed, and to acknowledge them. Errors -- even by renowned scientists -- occur, and history of physics teaches us: We progress by being wrong.

\section*{References}

%Create the reference section using BibTeX:

\end{document}